\documentclass[10pt,conference]{IEEEtran}
\IEEEoverridecommandlockouts

\usepackage{cite}
\usepackage{amsmath,amssymb,amsfonts}
\usepackage{algorithmic}
\usepackage{soul}
\usepackage{booktabs}
\usepackage{multirow}
\usepackage{float}
\usepackage{tcolorbox}
\usepackage{graphicx}
\usepackage{textcomp}
\usepackage{xcolor}
\usepackage{url}
\usepackage{tabularx} 
\usepackage{caption}
\usepackage{hyperref}

\hypersetup{
	pdfborder          = {0 0 0},
	breaklinks         = true,
	bookmarksnumbered  = true,
	pdfsubject         = {MSR 2021 Mining Challenge},
	pdfkeywords        = {Mining Software Repositories} {Unit Tests} {Bugs} {Simple Stupid Bugs} {Test Smells},
	pdftitle           = {On the Distribution of "Simple Stupid Bugs" in Unit Test Files: An Exploratory Study},
	pdfauthor          = {Anthony Peruma and Christian D. Newman}
}

\newcommand{\RQA}{RQ$_1$: To what extent do SStuBs occur in test files as they do in non-test files?}
\newcommand{\RQB}{RQ$_2$: To what extent do test files containing SStuB fixes also exhibit test smells?}

\begin{document}

\title{\Large On the Distribution of ``Simple Stupid Bugs'' in Unit Test Files: An Exploratory Study}

\author{\IEEEauthorblockN{Anthony Peruma}
\IEEEauthorblockA{\textit{Rochester Institute of Technology}\\
Rochester, New York, USA \\
axp6201@rit.edu}
\and
\IEEEauthorblockN{Christian D. Newman}
\IEEEauthorblockA{\textit{Rochester Institute of Technology}\\
Rochester, New York, USA \\
cnewman@se.rit.edu}
}

\maketitle

\begin{abstract}
A key aspect of ensuring the quality of a software system is the practice of unit testing. Through unit tests, developers verify the correctness of production source code, thereby verifying the system's intended behavior under test. However, unit test code is subject to issues, ranging from bugs in the code to poor test case design (i.e., test smells).
In this study, we compare and contrast the occurrences of a type of single-statement-bug-fix known as ``simple stupid bugs'' (SStuBs) in test and non-test (i.e., production) files in popular open-source Java Maven projects. Our results show that SStuBs occur more frequently in non-test files than in test files, with most fix-related code associated with assertion statements in test files. Further, most test files exhibiting SStuBs also exhibit test smells. We envision our findings enabling tool vendors to better support developers in improving the maintenance of test suites.
\end{abstract}


\section{Introduction}\label{Section:Introduction}
Unit testing is an essential strategy employed by developers to ensure the quality of their software system. Under this strategy, developers write code (i.e., test cases) that verifies the behavior of individual units of work of the system under test \cite{pressman2014software}. However, the test code written by developers is vulnerable to issues such as bugs (i.e., functional defects) and smells (i.e., bad programming practices), which impact not only the quality of the system but also the system's maintenance; specifically, the maintenance of test cases \cite{Bavota2012ICSM}.

However, correcting these software issues might not always be a complicated task, with some bugs being much easier to correct than others; often requiring a change to a single statement. These are sometimes referred to as ``simple stupid bugs'' (SStuBs) \cite{sstubsDataset}. While there exist studies that investigate defects in source code, these studies on program repair \cite{Gazzola2019TSE} do not differentiate defects occurring in test and non-test files, nor do these studies focus exclusively on SStuB-like defects. These studies mostly focus on a test case's ability to identify defects in the system under test \cite{Kim2013ICSE,Shamshiri2015ASE,Almasi2017ICSE}. Furthermore, studies that focus on the quality of test cases focus on test smells \cite{Garousi2018JSS} exhibited by test files such as investigating the flakiness (i.e., non-deterministic outcome) of test cases \cite{Spadini2018ICSME, grano2019scented}. In this paper, we focus on SStuBs in test files. We explore, compare, and contrast the existence of SStuBs in test files against non-test files. Furthermore, our study also looks at the co-occurrence between SStuBs and test smells.

\subsection{Goal \& Research Questions}
The goal of this study is to explore the quality of test suites from a functional and non-functional perspective, and the relationship between these viewpoints. Hence, we utilize SStuBs to represent the functional quality aspects of the test suite, and test smells for the non-functional quality aspects. As an exploratory study, we aim to have a better understanding of this problem and determine the feasibility of further research in this area. Our findings provide developers with insight to better design, implement, and maintain test suites. Additionally, our findings can improve the support provided by refactoring and other code quality tools. Hence, our study aims at answering the following research questions (RQs):
\begin{itemize}
    \item \textbf{\RQA}
    This RQ compares the occurrence of SStuBs in test files and non-test files. We examine the volume and types of SStuBs occurring in files along with the developers responsible for fixing SStuBs in the two groups of file types.
    
    
    \smallskip
    \item \textbf{\RQB}
    This RQ explores the existence of test smells in files containing fixes for SStuBs. This RQ aims to understand if SStuBs can be an indicator for the presence of test smells in the code.
\end{itemize}


\section{Data Extraction and Collection}\label{Section:DataExtraction}
\begin{figure}
 	\centering
 	\includegraphics[width=0.70\linewidth]{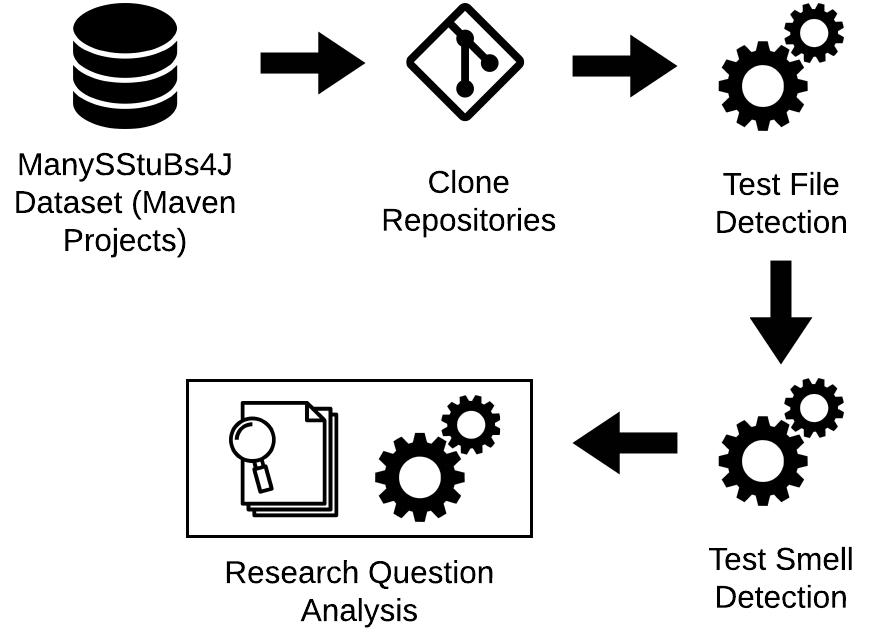}
 	\caption{Overview of our data extraction and collection process.}
 	\label{Figure:methodology}
\vspace{-0.65cm}
\end{figure}

\begin{table*}[!ht]
\fontsize{7.5}{10}\selectfont
\caption{Summary of the test smell detection rules utilized by \textsc{tsDetect}
\cite{Peruma2020FSE}.}
\vspace{-0.20cm}
\label{Table:smell_detection}
\resizebox{1\linewidth}{!}{%
 \begin{tabular}{ll} \hline
  \toprule
  \multicolumn{1}{c}{\textbf{Test Smell}} & \multicolumn{1}{c}{\textbf{Detection Rule}} \\ 
  \midrule
    Assertion Roulette & A test method contains more than one assertion statement without an explanation/message (parameter in the assertion method)  \\
    Conditional Test Logic & A test method that contains one or more control statements (i.e., if, switch, conditional expression, for, foreach and while statement)   \\
    Constructor Initialization & A test class that contains a constructor declaration\\
    Duplicate Assert & A test method that contains more than one assertion statement with the same parameters  \\ 
    Empty Test & A test method that does not contain a single executable statement\\
    Exception Handling & A test method that contains either a throw statement or a catch clause \\ 
    General Fixture & Not all fields instantiated within the \texttt{setUp} method of a test class are utilized by all test methods in the same test class \\
    Ignored Test & A test method or class that contains the \texttt{@Ignore} annotation \\ 
    Magic Number Test & An assertion method that contains a numeric literal as an argument \\ 
    Mystery Guest & A test method containing object instances of files and databases classes   \\
    Redundant Print & A test method that invokes either the \texttt{print} or \texttt{println} or \texttt{printf} or \texttt{write} method of the \texttt{System} class\\ 
    Redundant Assertion & A test method that contains an assertion statement in which the expected and actual parameters are the same \\ 
    Resource Optimism & A test method utilizes an instance of a \texttt{File} class without calling the method \texttt{exists()}, \texttt{isFile()} or \texttt{notExists()} methods of the object\\
    Sensitive Equality & A test method invokes the \texttt{toString()} method of an object\\
	Sleepy Test & A test method that invokes the \texttt{Thread.sleep()} method\\
    Unknown Test & A test method that does not contain a single assertion statement and \texttt{@Test(expected)} annotation parameter \\
  \bottomrule
  \end{tabular}
  }
 \vspace{-0.60cm}
\end{table*}

Figure \ref{Figure:methodology} shows a general overview of our data collection and extraction process. In the below subsections, we elaborate on the activities involved in the process. The dataset we utilize/generate is available at \cite{ProjectWebsite} for replication/extension.

\subsection{Source Dataset}
Our study utilizes the ManySStuBs4J dataset, which contains bug-fixing details of SStuBs extracted from popular open-source Java Maven projects \cite{sstubsDataset}. Contained within this dataset is the relative path of the file containing the fix, the commit id associated with the fix, and the type of SStuB fixed by the developer, among other details.

\subsection{Repository Cloning \& File Extraction}
In this stage of the process, we clone the Maven repositories that contain bug-fixing commits for SStuBs. The purpose of this activity is twofold. First, we extract metadata associated with the commit, such as the author (i.e., developer) of the fix. Second, we extract the source code files associated with each SStuB-fixing commit. The original Maven dataset contains 10,231 SStuB-fixing instances distributed over 84 projects. However, at the time of cloning these repositories, only 83 of the repositories were publicly available. Hence, in this study, we analyze 10,225 SStuB-fixing instances. 

\subsection{Test File Detection}
In this study, we only consider projects that utilize JUnit \cite{JUnitWebsite} as the testing framework since prior unit testing based research has frequently focused on JUnit (e.g., test smells \cite{Garousi2018JSS}).  JUnit recommends that developers either prepend or append the term ‘Test’ to the name of the production file to be tested (i.e., Test*.java or *Test.java). However, this mechanism can lead to false positives. For our study, we utilize the approach by \cite{Peruma2019CASCON} and \cite{Peruma2020FSE} to detect test files. In this approach, we utilize JavaParser \cite{JavaParserWebsite} to build an abstract syntax tree for each source file. We mark a file as a unit test file if the file contains JUnit import statements (i.e., \texttt{org.junit.*} or \texttt{junit.*}) and a test method. To be a test method, the method should have an annotation called \texttt{@Test} (JUnit 4), or the method name should start with `test' (JUnit 3). 

\subsection{Test Smell Detection}
For the detection of test smells we utilize \textsc{tsDetect} \cite{Peruma2020FSE}, an open-source test smells detection tool. \textsc{tsDetect} detects 19 test smell types and has been utilized in multiple studies \cite{Peruma2019CASCON,schvarcbacher2019investigating,Spadini2020Smells,Kim2020Smell,Peruma2020IWoR,Panichella2020ICSME}. Provided in Table \ref{Table:smell_detection} is a summary of the detection rules for each smell type. However, for this study, we ignore the \textit{Default Test} smell as this smell is exclusive to Android applications. We also ignore the smells \textit{Eager Test} and \textit{Lazy Test} as these smells require the class under test to be known; which is not known for this dataset. Hence, our analysis is limited to 16 smell types. Details about the smell types, along with examples, are available in \cite{Peruma2019CASCON}.

\subsection{Research Question Analysis}
To answer our research questions, we execute a series of database queries and custom scripts on the mined data. Additionally, where applicable, we provide examples from our dataset to supplement our findings. When reporting on our research questions, we first provide details about the experiment's methodology before presenting the findings.

\section{Experimental Results} \label{Section:Results}
\subsection{\RQA}
\noindent\textbf{Methodology.}
In this RQ, we group the files containing fixes for SStuBs into two groups-- test files and non-test files. The test files represent JUnit based unit testing files, while the non-test files represent the set of production files. We examine and compare details such as the distribution of SStuBs between the two groups, the developers responsible for applying the fix, and the clustering of file types in a bug-fixing commit.

\smallskip
\noindent\textbf{Results.}
First, we look at the projects containing SStuBs fixing commits. From the set of 83 projects in the dataset, 57 ($\approx$ 68.67\%) projects contain SStuBs occurring in both test and non-test files, while 25 ($\approx$ 30.12\%) projects contain SStuBs occurring in only non-test files, and only one project contains SStuBs occurring in only test files. Our next examination shows that developers apply SStuB fixes to 5,587 Java files. From this set, 1,066 ($\approx$ 19.08\%) are test files and 4,521 ($\approx$ 80.92\%) are non-test files. Moreover, from a total of 10,225 instances of SStuB fixes, 1,946 ($\approx$ 19.03\%) of these instances were in test files, while 8,279 ($\approx$ 80.97\%) were in non-test files. Furthermore, we observe, on average, 2.20 SStuBs occur in test files, and 2.37 occur in non-test files. 



Our next analysis focuses on SStuB categories. In total, there are 16 SStuB categories, the details of which are available at \cite{sstubsDataset}. In Table \ref{Table:BugType}, we provide the top five frequently occurring SStuB categories in test and non-test files. From this table, we observe that even though the categories are the same for both file types, the ratio of occurrences of these categories differs between file types. Further, we also observe that the categories \textit{Change Numeric Literal} and \textit{Change Modifier} occur at different positions relative to the other categories. These top five categories contribute to 88.23\% and 81.83\% of the complete set of occurrences in test and non-test file types, respectively. Additionally, we also observe that while the non-test files contain instances of all 16 categories, the test files do not exhibit the \textit{Change Operand} category. In general, we observe 314 ($\approx$ 16.13\%) SStuB instances in test files are associated with assertion statements. Examining the code statement containing the fix for \textit{Change Numeric Literal} in test files, we observe that 112 ($\approx$ 20.4\%) of the instances occur with an assertion statement (e.g., {\small$\texttt{assertEquals(2,map.size())} \rightarrow \texttt{assertEquals(3,map.size())}$} \cite{RQ2_00}). Furthermore, 116 ($\approx$ 21.13\%) instances are related to numeric values associated with time-related identifiers (e.g., {\small$\texttt{timeout=2000} \rightarrow \texttt{timeout=1000}$} \cite{RQ2_01}). In contrast, non-test files contain five instances of SStuBs in assertion statements and 31 time-related SStuBs. For cases under the \textit{Change Modifier} category, we observe that developers either remove or include the \texttt{static} keyword with a class or method in test files. We observe 14 cases falling under the \textit{Same Function More Args} category are related to assertion methods in test files. In most of these cases, developers include a textual message related to the assertion condition's failure (e.g., \cite{RQ2_02}). Additionally, we also encounter instances, in test files, where the methods that are updated are related to mocking; either API's related to Mockito or custom methods that accept mocked objects as arguments (e.g., {\small$\texttt{Mockito.any()} \rightarrow \texttt{Mockito.any(ProducerRecord.class)}$} \cite{RQ2_03}).

Our subsequent analysis is on the grouping of test and non-test files in a single commit. Since SStuBs can occur in both test and non-test files of a project, we are interested in knowing whether developers make fixes (i.e., commit) for both file types in unison or prioritize fixing one file type over another. Hence, for this analysis, we only focus on the 57 projects that contain SStuBs in test and non-test files. Our analysis shows that only one project meets this criterion. Furthermore, this project had only one commit operation that contained a test and non-test file. This phenomenon shows that developers prefer to treat SStuBs in isolation, most likely so that they can keep track of the changes and revert the change if necessary.

Finally, we examine the distribution of developers that fix SStuBs. Hence, we perform this analysis on the set of 57 projects containing test and non-test files. To detect unique developers, we utilize the same approach as \cite{Peruma2020JSS}. The approach utilizes the SStuB fixing commit author's email to identify unique developers for a project. In total, we encounter 1,116 unique bug fixing developers. From this set, 84 ($\approx$ 7.53\%) developers are responsible for fixing SStuBs occurring only in test files, while 789 ($\approx$ 70.70\%) fix SStuBs occurring only in non-test files. Developers that fix SStuBs in both test and non-test files account for 243 ($\approx$ 21.77\%). 




\begin{table}
\centering
\caption{Top five SStuB categories in (non-)test files.}
\vspace{-0.20cm}
\label{Table:BugType}
\begin{tabular}{@{}lrr@{}}
\toprule
\multicolumn{1}{c}{\textbf{SStuB Category}} & \multicolumn{1}{c}{\textbf{Count}} & \multicolumn{1}{c}{\textbf{Percentage}} \\ \midrule
\multicolumn{3}{c}{\textit{Test Files (Total Bug Type Instances: 1,946)}} \\
Change Identifier Used & 554 & 28.47\% \\
Change Numeric Literal & 549 & 28.21\% \\
Wrong Function Name & 310 & 15.93\% \\
Same Function More Args & 158 & 8.12\% \\
Change Modifier & 146 & 7.50\% \\
\multicolumn{3}{c}{\textit{Non-Test Files (Total Bug Type Instances: 8,279)}} \\
Change Identifier Used & 2,708 & 32.71\% \\
Change Modifier & 1,705 & 20.59\% \\
Wrong Function Name & 1,175 & 14.19\% \\
Same Function More Args & 600 & 7.25\% \\
Change Numeric Literal & 587 & 7.09\% \\ \bottomrule
\end{tabular}
\vspace{-0.60cm}
\end{table}

\smallskip
\noindent\textbf{Summary.}
This RQ shows that non-test files contain more SStuBs than test files. Additionally, we observe that the volume of SStuBs occurring in individual test and non-test files are very similar ($\approx$ 2 instances). We also observe that the majority of developers work on fixing SStuBs in non-test files than test files. Additionally, The top five popular SStuB categories for test and non-test files are the same. However, the ratio of occurrences of these categories differs. Furthermore, the code associated with SStuB fixes differs between test and non-test files; assertion statements in test files are frequently updated due to SStuB fixes.

\subsection{\RQB}
\noindent\textbf{Methodology.}
SStuBs represent the functional quality of a system. In contrast, test smells represent the non-functional quality of a system, both of which are important to the maintenance of the system's test suite. In this RQ, we look at the existence of test smells in test files exhibiting SStuBs. Additionally, we also investigate if the fixing of SStuBs causes a change in the number of test smell types exhibited by the file. To this extent, we utilize \textsc{tsDetect} to analyze the test files containing SStuB fixes and the prior version (i.e., commit) of the file for the existence of test smells.

\begin{table*}[!ht]
\centering
\caption{Co-occurrence between the top five frequently occurring SStuB bug categories and test smell types.}
\vspace{-0.20cm}
\label{Table:Co-occurrence}
\resizebox{1\linewidth}{!}{%
\begin{tabular}{@{}lrrrrrrrrrrrrrrrr@{}}
\toprule
\multicolumn{1}{c}{\textbf{\begin{tabular}[c]{@{}c@{}}SStuB \\ Category\end{tabular}}} & \multicolumn{1}{c}{\textbf{\begin{tabular}[c]{@{}c@{}}Assertion\\ Roulette\end{tabular}}} & \multicolumn{1}{c}{\textbf{\begin{tabular}[c]{@{}c@{}}Conditional\\ Test Logic\end{tabular}}} & \multicolumn{1}{c}{\textbf{\begin{tabular}[c]{@{}c@{}}Constructor\\ Initialization\end{tabular}}} & \multicolumn{1}{c}{\textbf{\begin{tabular}[c]{@{}c@{}}Empty\\ Test\end{tabular}}} & \multicolumn{1}{c}{\textbf{\begin{tabular}[c]{@{}c@{}}Exception\\ Handling\end{tabular}}} & \multicolumn{1}{c}{\textbf{\begin{tabular}[c]{@{}c@{}}General\\ Fixture\end{tabular}}} & \multicolumn{1}{c}{\textbf{\begin{tabular}[c]{@{}c@{}}Mystery\\ Guest\end{tabular}}} & \multicolumn{1}{c}{\textbf{\begin{tabular}[c]{@{}c@{}}Redundant\\ Print\end{tabular}}} & \multicolumn{1}{c}{\textbf{\begin{tabular}[c]{@{}c@{}}Redundant\\ Assertion\end{tabular}}} & \multicolumn{1}{c}{\textbf{\begin{tabular}[c]{@{}c@{}}Sensitive\\ Equality\end{tabular}}} & \multicolumn{1}{c}{\textbf{\begin{tabular}[c]{@{}c@{}}Sleepy\\ Test\end{tabular}}} & \multicolumn{1}{c}{\textbf{\begin{tabular}[c]{@{}c@{}}Duplicate\\ Assert\end{tabular}}} & \multicolumn{1}{c}{\textbf{\begin{tabular}[c]{@{}c@{}}Unknown\\ Test\end{tabular}}} & \multicolumn{1}{c}{\textbf{\begin{tabular}[c]{@{}c@{}}Ignored\\ Test\end{tabular}}} & \multicolumn{1}{c}{\textbf{\begin{tabular}[c]{@{}c@{}}Resource\\ Optimism\end{tabular}}} & \multicolumn{1}{c}{\textbf{\begin{tabular}[c]{@{}c@{}}Magic\\ Number Test\end{tabular}}} \\ \midrule
\textbf{Change Numeric Literal} & 375 & 373 & 354 & 326 & 375 & 366 & 344 & 346 & 304 & 355 & 364 & 372 & 372 & 368 & 355 & 373 \\
\textbf{Change Identifier Used} & 356 & 352 & 346 & 327 & 356 & 341 & 326 & 336 & 319 & 347 & 339 & 351 & 353 & 346 & 328 & 356 \\
\textbf{Wrong Function Name} & 192 & 189 & 183 & 170 & 192 & 177 & 170 & 177 & 160 & 184 & 178 & 188 & 190 & 183 & 172 & 192 \\
\textbf{Same Function More Args} & 110 & 110 & 109 & 97 & 110 & 106 & 104 & 102 & 100 & 109 & 105 & 110 & 110 & 107 & 104 & 110 \\
\textbf{Change Modifier} & 95 & 94 & 94 & 91 & 95 & 93 & 93 & 92 & 78 & 94 & 93 & 94 & 95 & 95 & 95 & 95 \\ 
\textit{\textbf{Other Categories}} & 164 & 163 & 162 & 155 & 164 & 161 & 158 & 157 & 153 & 162 & 159 & 163 & 164 & 163 & 159 & 164 \\ \bottomrule
\end{tabular} %
}
\vspace{-0.60cm}
\end{table*}

\smallskip
\noindent\textbf{Results.}
From the set of 1,066 test files containing SStuB fixes, 1,064 files show the presence of test smells, with each file exhibiting, on average, 15.41 smell types. Examining each smell type, we observe that each of the 16 smell types occur in over 80\% of the test files exhibiting SStuB fixes. Furthermore, the smell types \textit{Assertion Roulette} and \textit{Exception Handling} occur in all the test files. In terms of SStuB categories, we observe that \textit{Change Numeric Literal} and \textit{Change Identifier Used}  are the top two categories occurring in 375 and 356 smelly test file instances, respectively. This is in contrast to the popularity results shown in Table \ref{Table:BugType}.

Next, we look at the co-occurrence of each SStuB category with each smell type. In Table \ref{Table:Co-occurrence} we present the number of instances (i.e., test files) where a SStuB bug category co-occurs with a test smell. Due to space constraints, we only present the top five occurring SStuB bug categories. Our analysis shows that the smells \textit{Assertion Roulette}, \textit{Exception Handling}, and \textit{Magic Number Test} are the three smell types that most frequently co-occur with each of the SStuB categories. In contrast, the least co-occurring smell type is \textit{Redundant Assertion}.

Finally, we look at the version of the test file the developer modifies just before the commit containing the SStuB fix. Our objective is to determine if the smell count decreases when the developer fixes a SStuB. Our findings show that most SStuB fixes do not result in a decrease in smell count. More specifically, 853 ($\approx$ 80.17\%) instances do not show a change in smell count, while 182 ($\approx$ 17.11\%) instances show a decrease and 29 ($\approx$ 2.73\%) instances show an increase. Looking at the instances that show a decrease, we observe that, on average, 3.22, smell types are removed between the two versions of the file. We observe that the smells \textit{Assertion Roulette} and \textit{Exception Handling} do not show a reduction, while the smell \textit{Redundant Assertion} is a popular smell that reduces.

\smallskip
\noindent\textbf{Summary.}
Test files exhibiting SStuBs are very likely to contain test smells, with the \textit{Assertion Roulette} and \textit{Exception Handling} smell types frequently occurring in such files. Additionally, test files that exhibit certain types of SStuBs, such as \textit{Change Numeric Literal}, also exhibit test smells. However, developers rarely fix these smells when addressing SStuBs. 

\section{Discussion}\label{Section:Discussion}
As an exploratory study, our research aims to understand the extent to which SStuBs occur in test files and their relationship to test smell so as to provide direction to research areas that support developers in designing and maintaining test suites.

From \textbf{RQ$_1$}, we observe that SStuBs tend to occur more frequently in non-test files than test files. However, this should not signify that test files are of better quality than non-test files; this might be a situation where developers might not always be addressing these types of defects in test suites. It is interesting to note that developers usually address test and non-test SStuBs in separate commits. This is interesting since a change in code in a non-test file would require an appropriate change to the test code to ensure that the test case passes \cite{burns2013hudson}. Committing a non-test change independently of a test change would most usually cause the test case to fail in an automated build/test environment. Hence, it is most likely that such systems either do not use an automated build system or lack sufficient code coverage. Additionally, our observation of most developers working on fixing non-test SStuBs over test SStuBs can be a further indicator that developers prioritize non-test files over test files; and might need to adhere to a test-driven development approach \cite{beck2003test}. However, further research into this area is warranted to understand the rationale as to why developers treat test files differently from non-test files. Additionally, there needs to be research into tools that support developers with automatically finding/recommending changes to test files based on SStuB fixes applied to non-test files.

\textbf{RQ$_1$} also shows that the code related to SStuB fixes tend to differ between test and non-test files. We observe that the code usually associated with SStuB fixes in test files are frequently associated with assertion statements, such as in the case of the \textit{Change Numeric Literal} SStuB category. The frequent occurrence of issues related to assertion statements should not be surprising as asserts are one of the most fundamental parts of a test case. Furthermore, in \textbf{RQ$_2$}, we observe the frequent occurrence of the smell \textit{Assertion Roulette}; which corroborates with findings from prior test smell studies \cite{Bavota2012ICSM,Spadini2018ICSME}. These findings show that the occurrence of a specific SStuB category in the test suite indicates the occurrence of specific test smells in the same file, and thereby help developers in addressing multiple issues in the code that would otherwise be missed. A similar finding reported by Spadini et al. \cite{Spadini2018ICSME} shows an association between smelly tests and defect-proneness of the production code under test. For example, in our dataset, we observe instances where the \textit{Change Numeric Literal} SStuB category occurs due to changes made to the \texttt{Thread.sleep()} value, which is also an indicator of the \textit{Sleepy Test} smell (e.g., \cite{discussion_00}). This specific smell can lead to unexpected results as the processing time for a task differs when executed in various environments and configurations; most likely, the developer experienced this situation due to the sleep duration change. Lastly, our findings on the non-removal of smells also corroborate with research by Tufano et al. \cite{Tufano2017TSE}. Once more, all our findings can help tool/IDE vendors better equip their devices to better support developers to improve the functional and non-functional aspects of their test suites.

\section{Threats to Validity}\label{Section:Threats}
Even though the projects in our dataset are some of the most popular open-source Maven-based Java systems, the results may not generalize to systems written in other languages. Furthermore, we confine our analysis to the JUnit testing framework. However, prior unit testing-based research has frequently focused on JUnit \cite{Garousi2018JSS}. Our selection of \textsc{tsDetect} is due to its ability to detect multiple smell types and better detection performance \cite{Panichella2020ICSME}. Finally, even though non-test files have more SStuBs than test files, the per-file rate of SStuBs of test and non-test files are very similar. This aspect is interesting and requires more in-depth analysis, such as the possibility that SStuBs are not found at the same rate in test and non-test files.

\section{Conclusion \& Future Work}\label{Section:Conclusion}
In this study, we explore the occurrence of SStuBs in test files and their relation to test smells. We observe that SStuBs occur more frequently in non-test files than test files and that most of the fix related code differs between test and non-test files. Finally, we show that test files exhibiting SStuBs also exhibit test smells and tend to co-occur with specific types of SStuBs. These findings show that there is indeed scope for future research in this area, especially around the maintenance of test suites. Future work in this area includes investigating the flakiness of test suites brought about by SStuBs.

\bibliographystyle{ieeetr}
\bibliography{references}

\end{document}